\documentclass[%
superscriptaddress,
nofootinbib,
nobibnotes,
amsmath,amssymb,
aps,
floatfix,
]{revtex4-2}

\usepackage{placeins}
\usepackage{booktabs}
\usepackage{dcolumn}
\usepackage{array}
\usepackage{longtable}
\usepackage{float}
\usepackage{hyphenat}
\usepackage{natbib}
\usepackage{hyperref}

\usepackage{amsmath,amsfonts,amsthm,amssymb}
\usepackage{bm,graphicx,graphics,color}
\usepackage{epsf,epsfig}
\usepackage[all]{xy}
\newcommand*{\B}[1]{\ifmmode\bm{#1}\else\textbf{#1}\fi}

\def\bV{\mathbf{V}}

\def\no{\nonumber}
\def\lb{\label}
\def\be{\begin{equation}}
\def\ee#1{\label{#1}\end{equation}}
\newcommand{\ben}{\begin{eqnarray}}
\newcommand{\een}{\end{eqnarray}}

\usepackage{float}
\usepackage[english]{babel}

\usepackage[letterpaper,top=2cm,bottom=2cm,left=3cm,right=3cm,marginparwidth=1.75cm]{geometry}

\usepackage{amsmath}
\usepackage{graphicx}
\begin{document}

\title{Stellar structure model in the post-Newtonian approximation}

\author{Gilberto M. Kremer}
\email{kremer@fisica.ufpr.br}
\affiliation{Departamento de F\'{i}sica, Universidade Federal do Paran\'{a}, Curitiba 81531-980, Brazil}

\begin{abstract}

In this work the influence of the post-Newtonian corrections to the equations
of stellar structure is analysed. The post-Newtonian Lane-Emden equation follows from the corresponding momentum density balance equation. From a polytropic equation of state the solutions of the Lane-Endem equations in the
Newtonian and post-Newtonian theories are determined and the physical quantities for the \textit{Sun}, for the white dwarf \textit{Sirius
B} and for  neutron stars with masses $M\simeq1.4M_\odot,  1.8M_\odot$ and $2.0M_\odot$ are calculated. It is shown that the post-Newtonian corrections to the fields of mass density, pressure and temperature are negligible for the  \textit{Sun} and \textit{Sirius
B}, but for  stars with strong fields the differences become important. For the neutron stars analysed here the central pressure and the central temperature which follow from the post-Newtonian Lane-Emden equation are about fifty to sixty percent greater than those of the Newtonian  theory and the central mass density is about three to four percent smaller. 

\end{abstract}

\maketitle

%
\section{Introduction}           
\label{sect:intro}

The investigation of the internal structure of the stars is an old subject in the literature and this topic  was extensively described in the seminal books by Eddington \cite{Edd} and Chandrasekhar \cite{Chand}.

In astrophysics the Newtonian theory assumes  a prominent role in the characterization of the structure and dynamics  of stars, but also general relativity  assumes an important role in astrophysics. 

In the analysis of self-gravitating systems it is important to have an approximation scheme that provides a Newtonian description in the lowest order and relativistic effects as higher
order perturbations. For that end the  post-Newtonian theory can supply the desired  relativistic corrections to the Newtonian theory.  

The post-Newtonian theory was proposed by Einstein, Infeld and Hoffmann \cite{Eins} and refers to the solution of Einstein's field equations  from a method of successive approximations in the inverse power of the light speed (for a description of the method see e.g. the books \citep{Wein,Will,CF,GGKK}). The full Eulerian hydrodynamic equations in the first post-Newtonian approximation were derived by Chandrasekhar \cite{Ch1} and the corresponding ones in the second post-Newtonian approximation by Chandrasekhar and Nutku \cite{CNu}.

The post-Newtonian approximation is important in analyzing several problems: the equations of motion of binary pulsars \cite{r1,r2},
neutron stars \cite{n1,n2}, galaxy rotation curves \cite{c1,c2}, Jeans instability \cite{j1,j2,j3}, spherical accretion \cite{sp1}, stationary spherical self-gravitating systems \cite{sp2}, among others. 

In the last years the equations of stellar structure were analysed within the framework of the $f(R)$ theory where a modified Lane-Emden equation was derived  \citet{m1,m2,m3}.

The aim of this work is to investigate the  influence of the post-Newtonian corrections in the equations
of stellar structure which follow from the solution of the post-Newtonian Lane-Emden equation.   This equation is obtained  from the post-Newtonian momentum density balance equation for a stationary self-gravitating system where a polytropic equation of state is considered. The physical quantities  related with the mass density, pressure and temperature of a star are explicitly expressed in terms of the variables of the post-Newtonian Lane-Emden equation. From the polytropic solutions of the Lane-Endem equations in the Newtonian and post-Newtonian theories the physical quantities for the \textit{Sun},  the white dwarf \textit{Sirius
B} and for  neutron stars with masses $M\simeq1.4M_\odot,  1.8M_\odot$ and $2.0M_\odot$  are calculated. From the comparison of the Newtonian and post-Newtonian results for the physical quantities it  is shown that the post-Newtonian corrections to the fields of mass density, pressure and temperature are negligible for the  \textit{Sun} and \textit{Sirius
B}. However for  stars with strong fields the differences between  the two theories become important, since for the neutron stars analysed here the central pressure and the central temperature which follow from the post-Newtonian Lane-Emden equation are about fifty to sixty percent greater than those of the Newtonian  theory and the central mass density is about three to four percent smaller.

This paper is outlined as follows:
In Section \ref{sec2}, we introduce the post-Newtonian momentum density balance equation  and the corresponding Poisson equations. The post-Newtonian Lane-Emden equation is derived  in Section \ref{sec3}. In Section \ref{sec4}, we introduce the stellar structure equations in the post-Newtonian approximation. In Section \ref{sec5}, the numerical solutions for the mass density, pressure and
temperature   for the \textit{Sun}, \textit{Sirius B} and for the neutron stars are determined and the  Newtonian and post-Newtonian values for these fields are compared. Finally, in Section \ref{sec6}, we close the paper with
the conclusions.


\section{Post-Newtonian momentum density balance equation}\lb{sec2}

For a perfect fluid the energy-momentum tensor is given by 
\ben\lb{lane1}
T^{\mu\nu}=(\epsilon+p)\frac{U^\mu U^\nu}{c^2}+p g^{\mu\nu}.
\een
In the above equation  $p$ is the hydrostatic pressure, $U^\mu$ the four-velocity (such that $U^\mu U_\mu=c^2$),  $g^{\mu\nu}$ the metric tensor and $\epsilon$ the  energy density which has two contributions, one refers to the mass density $\rho c^2$ and another with its internal energy density $\varepsilon$, i.e, $\epsilon=\rho c^2\left(1+\varepsilon/c^2\right)$. Here we shall investigate a perfect fluid characterized by the polytropic equation of state $p=\kappa \rho^\gamma$, where $\kappa$ is a constant and $\gamma$ is related with the polytropic index  $n=1/(\gamma-1)$. For a polytropic fluid the internal energy density is given by $\varepsilon=p/[\rho(\gamma-1)]=np/\rho$.

In the derivation of the post-Newtonian approximations from Einstein's field equations in powers of the ratio $v/c$ -- where $v$ is a typical speed of the system and $c$ the light speed --   
the components of the metric tensor in the first post-Newtonian approximation read \cite{Ch1,GGKK}
\ben\lb{lane2a}
g_{00}=1-\frac{2U}{c^2}+\frac2{c^4}\left(U^2-2\Phi\right),\qquad
g_{0i}=\frac{\Pi_i}{c^3},\qquad g_{ij}=-\left(1+\frac{2U}{c^2}\right)\delta_{ij},
\een
where the Newtonian $U$, the scalar $\Phi$ and the vector $\Pi_i$ gravitational potentials     satisfy the Poisson equations  
\ben\lb{lane2b}
&&\nabla^2U=-4\pi G\rho,\qquad\nabla^2\Phi=-4\pi G\rho\left(V^2+U+\frac\varepsilon2+\frac{3p}{2\rho}\right),
\\
&&\nabla^2\Pi_i=-16\pi G\rho V_i+\frac{\partial^2U}{\partial t\partial x^i}.
\een
Here $\bV$ is the hydrodynamic three-velocity and $G$ the universal gravitational constant.

The  balance of the momentum density in the first post-Newtonian approximation obtained from the conservation of the energy-momentum tensor reads \cite{Ch1,GGKK}
\ben\no
\frac{\partial\sigma V_i}{\partial t}+\frac{\partial\sigma V_i V_j}{\partial x^j}+\frac{\partial}{\partial x^i}\left[p\left(1-\frac{2U}{c^2}\right)\right]
-\rho\frac{\partial U}{\partial x^i}\bigg[1+\frac1{c^2}\left(2V^2+\varepsilon-2U-\frac{p}\rho\right)\bigg]
\\\lb{lane3a}
-\frac\rho{c^2}V_j\left(\frac{\partial \Pi_i}{\partial x^j}-\frac{\partial \Pi_j}{\partial x^i}\right)+4\frac\rho{c^2}V_i\left(\frac{\partial U}{\partial t}+V_j\frac{\partial U}{\partial x^j}\right)
-\frac\rho{c^2}\left(2\frac{\partial \Phi}{\partial x^i}+\frac{\partial \Pi_i}{\partial t}\right)
=0,
\een
where $\sigma$ is the following abbreviation introduced by Chandrasekhar \cite{Ch1}
\ben\lb{lane3b}
\sigma =\rho\left[1+\frac1{c^2}\left(V^2+2U+\varepsilon+\frac{p}\rho\right)\right].
\een 

\section{Post-Newtonian Lane-Emden equation}\lb{sec3}
For the description of stellar structure models in the post-Newtonian approximation, we start with the balance equation of momentum density (\ref{lane3a}) by considering stationary self-gravitating systems where the hydrodynamic three-velocity vanishes, i.e, $\bf V = 0$.   Since in spherical coordinates the only dependence of the fields $\rho, p, U$ and $\Phi$  is on the radial  variable $r$, equation (\ref{lane3a}) becomes
\ben\lb{lane4a}
\left(1-\frac{2U}{c^2}\right)\frac{dp}{d r}
-\rho\frac{d U}{d r}\bigg[1+\frac1{c^2}\left(\varepsilon+\frac{p}\rho-2U\right)\bigg]-\frac{2\rho}{c^2}\frac{d \Phi}{d r}=0.
\een
By neglecting the $1/c^2$ terms the above equation reduces to the Newtonian limiting case $dp/dr=\rho dU/dr$. 

Equation (\ref{lane4a}) can be rewritten -- by taking into account that $\varepsilon=np/\rho$ and by considering terms up to $1/c^2$ -- as
\ben\lb{lane4b}
\frac1\rho\frac{dp}{d r}\left(1-\frac{n+1}{c^2}\frac{p}\rho\right)
-\frac{d}{d r}\left(U+2\frac\Phi{c^2}\right)=0.
\een

If we assume  the polytropic equation of state $p=\kappa \rho^{\frac{n+1}n}$, the differential equation (\ref{lane4b}) can be solved for  the mass density $\rho$ as function of the gravitational potentials $U,\Phi$, so that from the integration of the resulting equation we get
\ben\lb{lane5a} 
U+2\frac\Phi{c^2}=(n+1)\kappa\rho^{\frac1n}\left(1-\frac{\kappa(1+n)}{2c^2}\rho^{\frac1n}\right).
\een
In the above equation it was considered that the gravitational potentials $U$ and $\Phi$ and the mass density $\rho$ vanish at the boundary of the star. The argument that $U$ vanish at the boundary is due to Eddington \cite{Edd}, here we extend it to the post-Newtonian gravitational potential $\Phi$.

We   can  solve (\ref{lane5a}) for $\rho$ up to order $1/c^2$, yielding
\ben\lb{lane5b}
\rho=\left[\frac{U+\frac{2\Phi}{c^2}}{(n+1)\kappa\left(1-\frac{\kappa(1+n)}{2c^2}\rho^{\frac1n}\right)}\right]^n\approx\left(\frac{U}{(n+1)\kappa}\right)^n\left[1+\frac{n}{c^2}\left(\frac{U}2+\frac{2\Phi}U\right)\right].
\een

 The Poisson equations (\ref{lane2b}) for the gravitational potentials  $U$ and $\Phi$ in spherical coordinates, for stationary systems ruled by a polytropic equation of state $p=\kappa \rho^{\frac{n+1}n}$ and $\varepsilon=np/\rho$ become
\ben\lb{lane6a}
\frac1{r^2}\frac{d}{dr}\left(r^2\frac{dU}{dr}\right)=-4\pi G\rho,
\qquad
\frac1{r^2}\frac{d}{dr}\left(r^2\frac{d\Phi}{dr}\right)=-4\pi G\rho\left(U+\frac{3+n}{2}\kappa\rho^\frac1n\right).
\een

The combination of the two Poisson equations (\ref{lane6a}) yields
\ben\lb{lane6b}
\frac1{r^2}\frac{d}{dr}\left[r^2\frac{d}{dr}\left(U+2\frac\Phi{c^2}\right)\right]
=-4\pi G\rho\left[1+\frac2{c^2}\left(U+\frac{3+n}2\kappa\rho^\frac1n\right)\right].
\een

The elimination of the potentials $U,\Phi$ from (\ref{lane6b})  by   using (\ref{lane5a})  results the following  differential equation for the mass density 
\ben\lb{lane7a}
\kappa(n+1)\frac1{r^2}\frac{d}{dr}\left[r^2\frac{d}{dr}\left(\rho^{\frac1n}-\frac{1+n}{2c^2}\kappa\rho^{\frac2n}\right)\right]
 =-4\pi G\rho\left[1+\frac{5+3n}{c^2}\kappa\rho^{\frac1n}\right].
\een

The dimensionless Lane-Emden equation is obtained from the introduction of the dimensionless variables \cite{Edd,Chand}
\ben\lb{lane7b}
z=\frac{r}a,\qquad u(z)=\left(\frac\rho{\rho_c}\right)^\frac1n,\qquad
a=\sqrt{\frac{(n+1)\kappa}{4\pi G}\rho_c^{\frac{1-n}n}},
\een
where $\rho_c$ denotes the mass density at the center of the star.

The introduction of the new variables (\ref{lane7b}) into (\ref{lane7a}) leads to the Lane-Emden equation in the first post-Newtonian approximation
\ben\no
&&\bigg(1-\frac{(1+n)p_c}{c^2\rho_c}u(z)\bigg)\bigg[\frac{d^2u(z)}{dz^2}+\frac2z\frac{du(z)}{dz}\bigg]
-\frac{(1+n)p_c}{c^2\rho_c}\bigg(\frac{du(z)}{dz}\bigg)^2
\\\lb{lane8a}
&&\qquad=-u(z)^n\bigg(1+\frac{(5+3n)p_c}{c^2\rho_c}u(z)\bigg),
\een
where $p_c=\kappa\rho_c^\frac{n+1}n$ is the hydrostatic pressure at the center of the star.

An equivalent version of the first post-Newtonian approximation of the Lane-Emden equation is obtained from the multiplication of (\ref{lane8a})  by $[1+{(5+3n)p_cu(z)}/{c^2\rho_c}]$ and considering   terms up to the order $1/c^2$, yielding
\be
\bigg(1-\frac{(6+4n)p_c}{c^2\rho_c}u(z)\bigg)\bigg[\frac{d^2u(z)}{dz^2}+\frac2z\frac{du(z)}{dz}\bigg]
-\frac{(1+n)p_c}{c^2\rho_c}\bigg(\frac{du(z)}{dz}\bigg)^2+u(z)^n=0.
\ee{lane8b}
If in the above equation we do not consider the $1/c^2$--terms  the Newtonian limit of the Lane-Emden equation  is recovered, namely
\ben\lb{lane8c}
\frac1{z^2}\frac{d}{dz}\left(z^2\frac{du(z)}{dz}\right)=-u(z)^n.
\een
Furthermore, by considering the perfect fluid equation of state for the hydrostatic pressure at the center of the star $p_c=\rho_c kT_c/m=\rho_ckT_c/\mu m_\mu$ -- where $T_c$  represents the temperature at the star center, $\mu$ the mean molecular weight and $m_\mu$ the unified atomic mass -- we can write
\ben\lb{lane8d}
\frac{p_c}{\rho_cc^2}=\frac{kT_c}{mc^2}=\frac{kT_c}{\mu m_\mu c^2}.
\een
Note  that ${p_c}/{\rho_cc^2}$ represents the ratio of the thermal energy of the fluid at the star center $kT_c$ and the rest energy of its particles $mc^2$.

In astrophysics, the Lane-Emden equation is used to describe thermodynamic system structures characterized by polytropic fluids, considering the gravitational interaction. This equation allows us to determine some physical quantities for these systems, such as pressure, density, and temperature. 
	
\section{Physical quantities of stars}\lb{sec4}

In this section we follow  Eddington \cite{Edd} and Chandrasekhar \cite{Chand} and give the expressions for the mass, radius, pressure, mass density and temperature of the stars which follow from the Lane-Emden equation.

The Lane-Emden equation (\ref{lane8b}) will be solved by considering  the boundary conditions  
\ben\lb{lane9}
u(0)=1,\qquad \frac{du(z)}{dz}\bigg\vert_{z=0}=0.
\een

The numerical solution of (\ref{lane8b}) represents a monotonically decreasing behavior of $u(z)$ and  its first zero --  denoted by
$z\vert_{u=0}=R_*$ -- corresponds to the surface of the star. From (\ref{lane7b}) the radius of the star becomes
\ben\lb{lane10}
R=aR_*=\sqrt{\frac{(n+1)\kappa}{4\pi G}\rho_c^\frac{1-n}n}R_*.
\een

For a sphere with radius $R$ its inner mass $M(R)$ is given by
\ben\lb{lane11a}
M(R)=\int_0^{R}4\pi \sqrt{\gamma_*}\rho r^2dr,
\een
Here $\gamma_*$ denotes the determinant of the spatial metric tensor, which  by considering terms up to $1/c^2$ order reads
\ben\lb{lane11b}
\sqrt{\gamma_*}=\sqrt{\frac{-g}{g_{00}}}=\left(1+\frac{3U}{c^2}\right)=\left(1+\frac{3(n+1)\kappa\rho^{\frac1n}}{c^2}\right)
=\left(1+\frac{3(n+1)p_c}{c^2\rho_c}u(z)\right),
\een
by taking into account (\ref{lane5a}),  (\ref{lane7b}) and the expression for the determinant of the metric tensor in the first post-Newtonian approximation $g=-(1+4U/c^2)$. 

The mass of the star which follows from the Lane-Emden equation (\ref{lane8b}) is given by 
 \ben\no
&&M(R)=4\pi a^3\rho_c\int_0^{R_*}\left(1+\frac{3(n+1)p_c}{c^2\rho_c}u(z)\right) z^2 u^n dz
\\\no
&&\qquad=-4\pi a^3\rho_c\int_0^{R_*}\bigg\{\bigg(1-\frac{(3+n)p_c}{c^2\rho_c}u(z)\bigg)\bigg[\frac{d^2u(z)}{dz^2}+\frac2z\frac{du(z)}{dz}\bigg]
\\\lb{lane11c}
&&\qquad-\frac{(1+n)p_c}{c^2\rho_c}\bigg(\frac{du(z)}{dz}\bigg)^2\bigg\}z^2dz=4\pi\rho_c a^3 M_*.
\een
In the second equality above we have considered only terms up to the $1/c^2$ order.

From the elimination of $a$ and $\rho_c$ from (\ref{lane11c})  by using  (\ref{lane7b}) and (\ref{lane10}) we get that the mass of the star becomes
\ben\lb{lane12a}
M(R)=4\pi\left[\frac{(n+1)\kappa}{4\pi G}\right]^\frac{n}{n-1}\left(\frac{R}{R_*}\right)^\frac{n-3}{n-1}M_*.
\een

Now we can build the  mass-radius relationships by taking into account (\ref{lane10}) and (\ref{lane12a}), yielding
\ben\lb{lane12b}
\frac{GM(R)}{M_*}\frac{R_*}{R}=(n+1)\kappa \rho_c^\frac1n,
\qquad\left(\frac{GM(R)}{M_*}\right)^{n-1}\left(\frac{R_*}R\right)^{n-3}=\frac{[(n+1)\kappa]^n}{4\pi G}.
\een

The quantities $R_*$ and $M_*$ can be determined from the Lane-Emden equation (\ref{lane8b}) once the  mass $M(R)$ and radius $R$ of a star  are known. Furthermore,  for fixed values of the polytropic index $n$, the values of $\kappa$ and $\rho_c$   follow from (\ref{lane12b}).

We may also express   the central mass density of the star as function of the mean mass density of the star $\overline\rho$, namely 
\ben\lb{lane13a}
\overline\rho=\frac{M(R)}{4\pi R^3/3},\qquad\hbox{hence}\qquad\rho_c=\frac{R_*^3}{3M_*}\overline\rho,
\een
thanks to (\ref{lane10}) and (\ref{lane11c}).

From the polytropic equation of state $p_c=\kappa\rho_c^\frac{1+n}n$ together with (\ref{lane12b}) and (\ref{lane13a}) we can determine the central pressure of the star
\ben\lb{lane13b}
p_c=\frac{GM(R)}{M_*}\frac{R_*}{R}\frac{\rho_c}{n+1}=\frac{GM(R)}{M_{*}}\frac{R_{*}}{R}\frac{\overline\rho}{n+1}\frac{R_{*}^3}{3M_{*}},
\een
furthermore, from the equation of state of a perfect fluid we get  the temperature at the center of the star  
\ben\lb{lane13c}
T_c=\frac{\mu m_\mu}k\frac{p_c}{\rho_c}= \frac{\mu m_\mu}{k(n+1)}\frac{GM(R)}{M_*}\frac{R_*}{R}.
\een

The mass density, pressure and temperature as functions of the dimensionless radial distance $z$  follows from the polytropic equation of state and  (\ref{lane7b}), yielding
\ben\lb{lane13d}
\rho(z)=\rho_c \,u(z)^n,\qquad p(z)=p_c \,u(z)^{n+1}, \qquad T(z)=T_c\,u(z).
\een

\section{Polytropic solutions of the Lane-Emden equation}\lb{sec5}

A star is identified as a self-gravitating spherically symmetrical mass of a highly ionized gas at equilibrium which is held together by its own gravity. Normally a star is considered to be  composed  by three kinds of species:  hydrogen, helium and heavy elements, which for the purpose of the calculations are not specified.

If $X$, $Y$ and $Z$ denote the mass fraction of hydrogen, helium and heavy elements, respectively, for a mixture with these three species we must have that $X+Y+Z=1$ and the mean molecular weight becomes \cite{Chand}
\ben
\mu=\frac1{2X+3Y/4+Z/2}=\frac4{2+6X+Y}.
\een

In this work we are interested in determining the influence of the post-Newtonian approximation in the  stellar structures: neutron stars, white  dwarfs, and the Sun.  Neutron stars are formed from a  gravitational collapse of massive stars at the end of their life and practically  have only neutrons so that $\mu=1$.  The mass fractions for the Sun are $X=0.73$, $Y=0.25$ and $Z=0.02$ \cite{Sun} and its mean molecular weight is $\mu=0.6$.
White dwarfs are compact objects with low luminosity  and here we shall investigate  the white dwarf \textit{Sirius B} --  which is the companion that orbits around the \textit{Sirius}  star --  where there exists almost heavy metals $Z\approx1$,  are devoid of hydrogen and helium so that $X=Y\approx0$  so that  the mean molecular weight is $\mu=2$.

The \textit{Sun} has a radius $R_\odot=6.96\times 10^8$m, a mass  $M_\odot=1.989\times 10^{30}$kg  and the polytropic index usually adopted for it is $n=3$. 
For  white dwarf stars with higher masses
the polytropic index can also be considered  as $n=3$ and the  \textit{Sirius B} has mass $M=1.5 M_\odot$ and radius  $R=8.4\times 10^{-3}R_\odot$. 

Neutron stars are represented by an equation of state with a polytropic index  $n\simeq1$ \cite{eos} and we will focus our attention to  neutron stars    with masses $M\simeq1.4, M_\odot,  1.8 M_\odot$ and $2.0 M_\odot$.  According to \cite{mass1,mass} the  radii of the neutron stars are in the range 8.3 km $\leq$ R $\leq$ 12 km   for all neutron stars. Here we adopted the following radii for the neutron stars:  $R\simeq9.8$ km for $M\simeq1.4 M_\odot$, $R\simeq9$ km for $M\simeq1.8 M_\odot$ and $R\simeq8.7$ km for $M\simeq2.0 M_\odot$. The radius of  the neutron star  corresponding to the mass $M\simeq1.8 M_\odot$ was taken as $R\simeq9$ km  and the radii of the neutron stars with masses $1.4 M_\odot$ and $2.0 M_\odot$ were obtained by using the relationship $R\propto M^{-\frac13}$.

First we analyze  the results that follow from the Newtonian Lane-Emden equation  for the  \textit{Sun},
 \textit{Sirius B} and the neutron stars. In Table \ref{tab5.1} the first zeros were found as numerical solutions of the Newtonian   Lane-Emden equation (\ref{lane8c}) and the mean and central mass densities, central pressure and central temperature were calculated from (\ref{lane13a}), (\ref{lane13b}) and (\ref{lane13c}) when the post-Newtonian correction $p_c/c^2\rho_c$ is not considered.   The polytropic indexes adopted are: $n=1$ for the  neutron stars and $n=3$ for the  \textit{Sun} and  \textit{Sirius B} .  We infer from this table that  the \textit{Sun} and  \textit{Sirius B} have the same first zeros,  since they have the same polytropic index.  Furthermore,  the values of the central quantities for the  neutron stars  are several  orders of magnitude  greater  than  those of the white dwarf \textit{Sirius B} and the same occurs when we compare the values of the central quantities of the latter with those   of the \textit{Sun}. This behavior follows from the fact that smaller radius and a greater mass lead to an increase in the values of the central quantities.
  
\begin{table}[ht]
\caption{First zeros, central and mean  mass densities, central pressures and central temperatures calculated from the Newtonian Lane-Emden equation (\ref{lane8c}).}
{\begin{tabular}{|c|c|c|c|c|c|c|}
  \hline
  &$R_*$&$M_*$&$\overline\rho$ (kg/m$^3$)&$\rho_c$ (kg/m$^3$)&$p_c$ (Pa)&$T_c$ (K)\\
  \hline
  \textit{Sun}&6.90&2.02&$1.41\times10^3$&$7.64\times10^4$ &$1.25\times10^{16}$ &$1.18\times10^7$ \\
    \hline
     \textit{Sirius B}&6.90&2.02&$2.89\times10^9$&$1.56\times10^{11}$ &$3.34\times10^{24}$ &$5.14\times10^9$ \\
  \hline
  $1.4M_\odot$&3.14&3.14&$7.06\times10^{17}$&$2.33\times10^{18}$ &$2.21\times10^{34}$ &$1.14\times10^{12}$ \\\hline
 $1.8M_\odot$&3.14&3.14&$1.17\times10^{18}$&$3.87\times10^{18}$ &$5.14\times10^{34}$ &$1.59\times10^{12}$ \\\hline
 $2.0M_\odot$&3.14&3.14&$1.44\times10^{18}$&$4.76\times10^{18}$ &$7.27\times10^{34}$ &$1.84\times10^{12}$ \\
  \hline
\end{tabular}
\lb{tab5.1}}
\end{table}

From the comparison of the Lane-Emden equations in the post-Newtonian (\ref{lane8b}) and  Newtonian (\ref{lane8c})  theories we note that the difference between them  lies on the terms that are multiplied by  $p_c/\rho_c c^2=kT_c/mc^2$, which  corresponds to the ratio of the thermal energy of the fluid at the star center $kT_c$ and the rest energy of its particles $mc^2=\mu m_\mu c^2$. This parameter was determined from the values of the central temperature $T_c$ given in Table \ref{tab5.1} and  are shown in Table \ref{tb}.
\begin{table}[ht]
\caption{Values of the ratio $p_c/\rho_c c^2=kT_c/mc^2$.}
{\begin{tabular}{|c|c|c|c|c|c|}
  \hline
  &\textit{Sun}&\textit{Sirius B}&$1.4M_\odot$&$1.8M_\odot$&$2.0M_\odot$\\
  \hline
$kT_c/mc^2$  &$1.19\times 10^{-6}$&$2.37\times 10^{-4}$ &$1.05\times 10^{-1}$&$1.48\times 10^{-1}$&$1.70\times 10^{-1}$ \\
    \hline
    \end{tabular}
\lb{tb}}
\end{table}

We may conclude from the Table \ref{tb} that the values of the ratio $p_c/\rho_c c^2=kT_c/mc^2$ for the \textit{Sun} and \textit{Sirius B} are very small so that the post-Newtonian corrections to the Lane-Emden equation are negligible and  the values given in Table \ref{tab5.1} for these stars remain practically unchanged. 

\begin{table}[ht]
\caption{First zero, central and mean mass densities, central pressure and central temperature from the post-Newtonian Lane-Emden equation (\ref{lane8b}) for the neutron stars.}
{\begin{tabular}{|c|c|c|c|c|c|c|}
  \hline
  &$R_*$&$M_*$&$\overline\rho$ (kg/m$^3$)&$\rho_c$ (kg/m$^3$)&$p_c$ (Pa)&$T_c$ (K)\\
    \hline
  $1.4M_\odot$&2.56&1.75&$7.06\times10^{17}$&$2.26\times10^{18}$ &$3.14\times10^{34}$ &$1.67\times10^{12}$ \\
 \hline
  $1.8M_\odot$&2.43&1.52&$1.17\times10^{18}$&$3.70\times10^{18}$ &$7.86\times10^{34}$ &$2.56\times10^{12}$ \\
 \hline
  $2.0M_\odot$&2.38&1.43&$1.44\times10^{18}$&$4.55\times10^{18}$ &$1.16\times10^{35}$ &$3.06\times10^{12}$ \\
  \hline
\end{tabular}}
\lb{tab5.2}
\end{table}

\begin{figure}[h]
\centerline{\includegraphics[width=10cm]{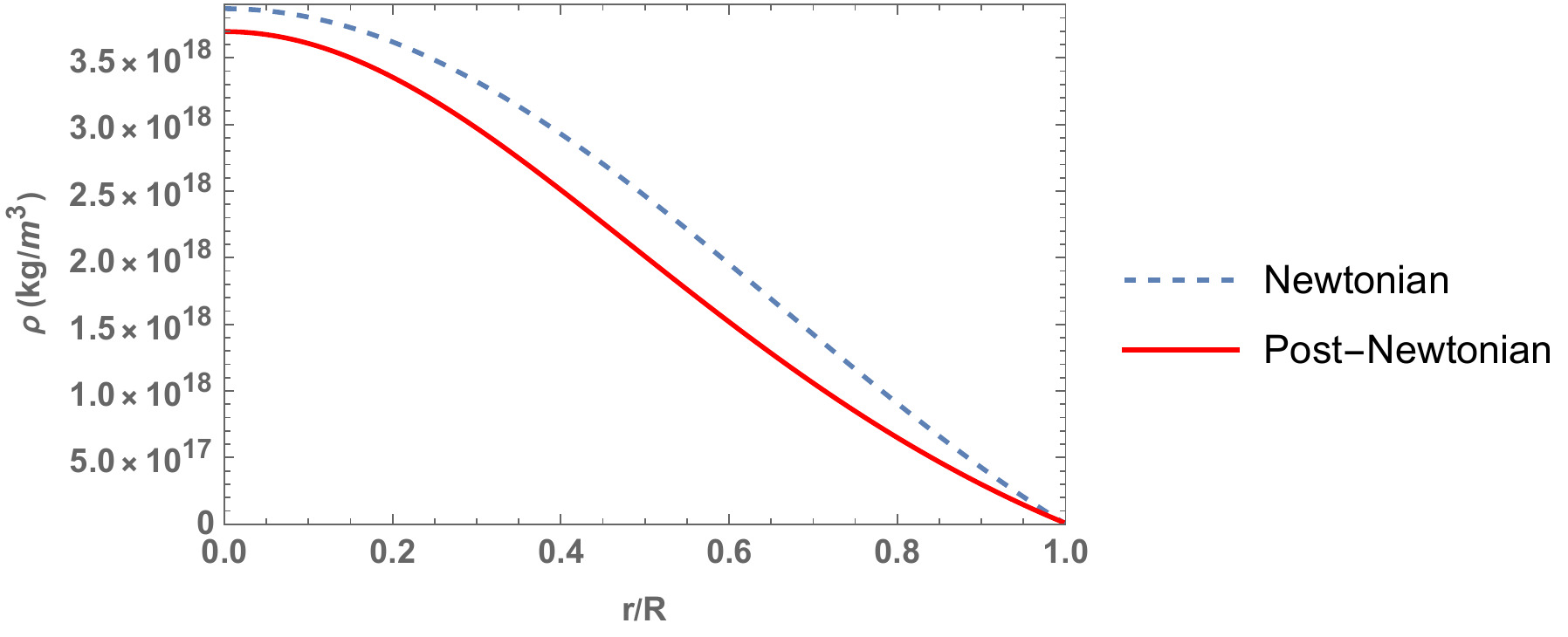}}\caption{Mass densities $\rho$ as functions of the normalized radius $r/R$ for the neutron star $1.8M_\odot$. Solid line -- post-Newtonian solution, dashed line -- Newtonian solution.}\lb{fig0}
\end{figure}

The post-Newtonian corrections are important for more massive stars like the neutron stars, since  its central temperature is at least three orders of magnitude greater than those of the \textit{Sun} and \textit{Sirius B} and the  ratio of the thermal energy at the star center  and the rest energy of the particle is $kT_c/mc^2\approx 10^{-1}$. In Table \ref{tab5.2} the first zero and the values for the central quantities -- calculated from the post-Newtonian Lane-Emden equation (\ref{lane8b}) -- are given for   the neutron stars.  We may infer from the comparison of the values  for   the neutron stars  given in the Tables \ref{tab5.1} and  \ref{tab5.2}  that in the post-Newtonian theory the values  for the central pressure and temperature are about fifty to sixty percent larger than those of the Newtonian theory, while the value for the central mass density is about  three to four percent smaller.

\begin{figure}[h]
\centerline{\includegraphics[width=10cm]{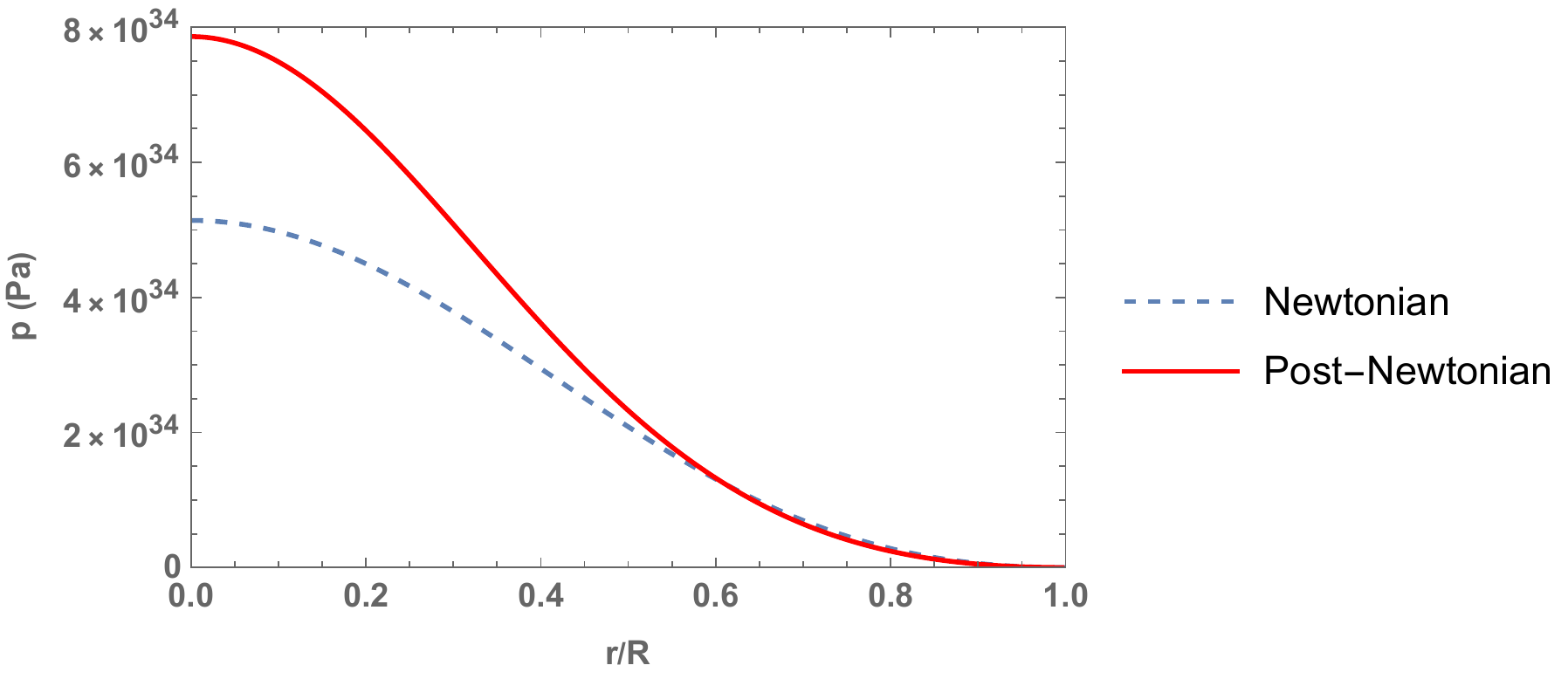}}\caption{Pressures $p$ as functions of the normalized radius $r/R$ for the neutron star $1.8M_\odot$.  Solid line -- post-Newtonian solution, dashed line -- Newtonian solution. }\lb{fig1}
\end{figure}

\begin{figure}[H]
\centerline{\includegraphics[width=10cm]{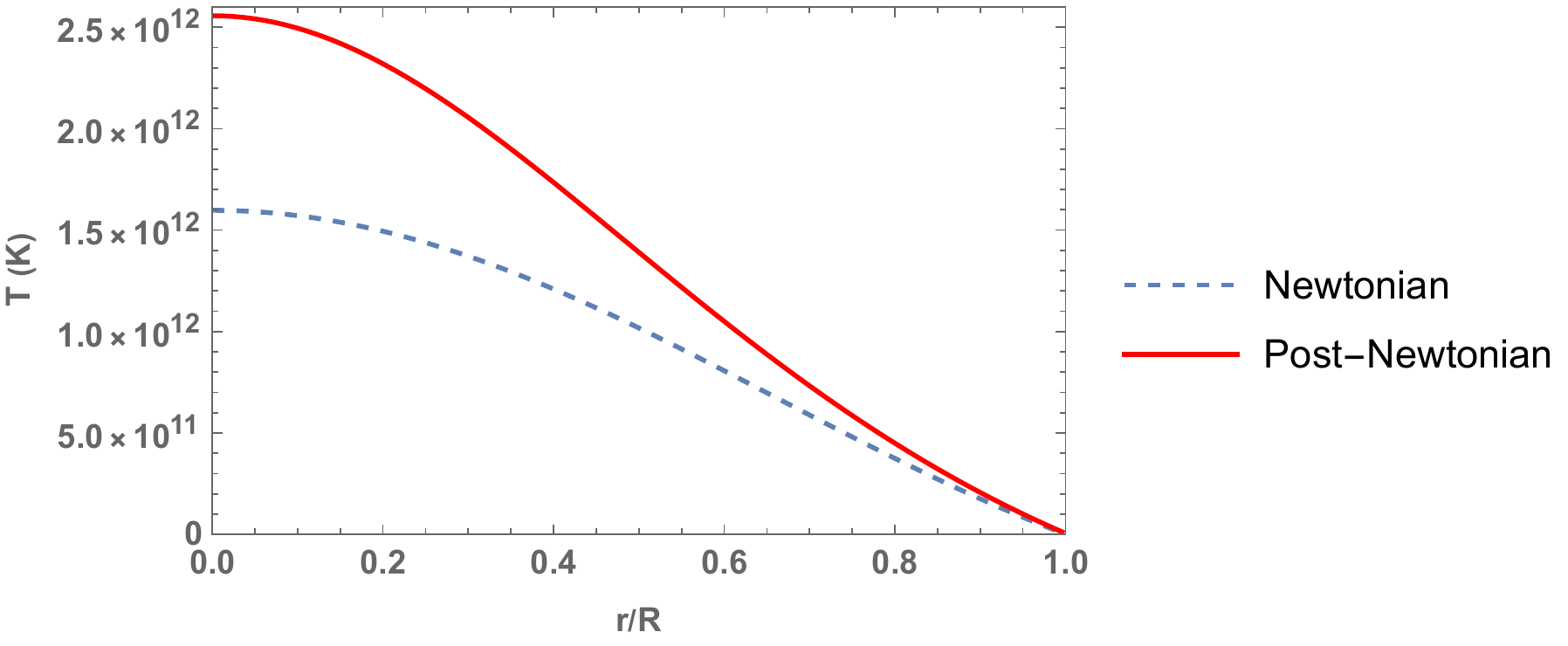}}\caption{Temperatures $T$ as functions of the normalized radius $r/R$ for the neutron star $1.8M_\odot$.   Solid line -- post-Newtonian solution, dashed line -- Newtonian solution.}\lb{fig2}
\end{figure}

From the knowledge of the numerical solutions which follow from the Newtonian and post-Newtonian Lane-Emden equations for $u(z)$ and of the central quantities for $\rho_c, p_c$ and $T_c$, one may obtain from (\ref{lane13d}) the   behaviors of the mass density $\rho$, pressure $p$ and temperature $T$ as functions of the normalized radius $r/R$.  In Figure \ref{fig0} the mass density $\rho$ for the neutron star with mass $1.8M_\odot$ is plotted as a function normalised radius $r/R$, while the Figures \ref{fig1} and \ref{fig2} represent the pressure $p$ and the temperature $T$, respectively. While the post-Newtonian solutions for the pressure and temperature are greater than those of the Newtonian ones, the Newtonian solution for the mass density is greater than the post-Newtonian solution.
All three plots show that all fields have a monotonically decreasing behavior with respect to the normalized radius.

The value of the mass density at the crust can be obtained from the limiting value when $r/R\rightarrow1$ and its value is of order $10^{15}$, while from Figure 1 we infer that the mass density value at the center of the neutron star is of order $10^{18}$. Both values are  one magnitude order greater  than those reported in the literature. Note that here a polytropic equation of state was assumed and there are other equations of state that were proposed in the literature to describe properly the neutron stars \cite{hh}.

\section{Conclusions}\lb{sec6}

The aim of this work was to analyse the influence of the post-Newtonian corrections in the stellar  structure equations. Starting from the post-Newtonian momentum density balance equation,  the corresponding Lane-Emden equation was obtained.   By assuming a polytropic equation of state, the solutions of the Lane-Endem equations in the Newtonian and post-Newtonian theories were determined. The physical quantities for the \textit{Sun}, for the white dwarf \textit{Sirius B} and for neutron stars with masses $M\simeq1.4 M_\odot,  1.8 M_\odot$ and $2.0 M_\odot$ were numerically calculated by considering the Newtonian and post-Newtonian solutions of the Lane-Emden equations. It was shown that the post-Newtonian corrections were negligible for the \textit{Sun} and for \textit{Sirius B}. For stars with strong fields the post-Newtonian corrections become important, so that for  the neutron stars analysed here the central pressure and the temperature which follow from the post-Newtonian Lane-Emden equation are about fifty to sixty percent greater than those of the Newtonian  one and the central mass density is about three to four percent smaller. 

\begin{acknowledgements}
This work was supported by Conselho Nacional de Desenvolvimento Cient\'{i}fico e Tecnol\'{o}gico (CNPq), grant No.  304054/2019-4.
\end{acknowledgements}



\end{document}